\documentclass[12pt]{article} 

\usepackage{epsfig}

\def\en{\epsilon}
\def\Xv{{\bf X}}

\def\rv{{\bf r}}
\def\vv{{\bf v}}

\def\bv{{\bf b}}

\def\pv{{\bf p}}

\def\kv{{\bf k}}

\def\Ev{{\bf E}}

\def\bcl{b_{\rm cl}}

\def\bv{{\bf b}}

\def\deltare{\delta{\bf r}}

\def\LLm{l_{-1}}
\def\LL{l_0}
\def\LLp{l_2}

\def\RB{R}

\def\BBm{b_{-2}}
\def\BB{b_0}
\def\BBp{b_1}

\def\Airi{{\cal A}}

\def\lambdabar{\lambda\raise0.4ex\hbox{\kern-0.5em\hbox{--}}\ }
\def\lambdaC{\lambda\raise0.5ex\hbox{\kern-0.5em\hbox{--}}_{\rm C}}
\def\lambdabarc{\lambda\raise0.5ex\hbox{\kern-0.5em\hbox{--}}_{\rm c}}
\def\lesssim{{\lower0.5ex\hbox{$\stackrel{<}{\sim}$}}}
\def\gtrsim{{\lower0.5ex\hbox{$\stackrel{>}{\sim}$}}}

\begin{document}

\centerline{
IMPACT PARAMETER PROFILE \\
OF SYNCHROTRON RADIATION
}

\bigskip
\medskip

\centerline{Xavier ARTRU}

\bigskip
\medskip

\centerline{Institut de Physique Nucl\'eaire de Lyon,}
\centerline{Universit\'e Claude-Bernard \& IN2P3-CNRS,}
\centerline{69622 Villeurbanne, France. %
{\it Email: x.artru@ipnl.in2p3.fr}
}
\begin{abstract}

The horizontal impact parameter profile of synchrotron radiation, for fixed vertical 
angle of the photon, is calculated. This profile is observed through an astigmatic 
optical system, horizontally focused on the electron trajectory
and vertically focused at infinity. It is the product of the usual angular distribution
of synchrotron radiation, which depends on the vertical angle $\psi$, 
and the profile function of a caustic staying at distance 
$\bcl = (\gamma^{-2} + \psi^2) \RB/2 $ 
from the orbit circle, $\RB$ being the bending radius and $\gamma$ the Lorentz factor. 
The {\it classical impact parameter} $\bcl$ is connected to the Schott term of 
radiation damping theory. 
The caustic profile function is an Airy function squared. 
Its fast oscillations allow a precise determination of the horizontal beam width.

\end{abstract}%

\medskip
\noindent
keywords:
{\it synchrotron radiation, beam diagnostics, radiation damping.}


\section{Theoretical starting point}

Classically \cite{Bignon,Qasmi}, a photon from synchrotron radiation is emitted not exactly from the electron orbit 
but at some distance or {\it impact parameter} 
\begin{equation}
\bcl \ \equiv \ R_{phot} - \RB 
\ = \ {\RB \over v \cos\psi} - \RB \ \simeq \ {\gamma^{-2} + \psi^2 \over2} \, \RB
\,,
\label{bcl}
\end{equation}
from the cylinder which contains the orbit.
$\RB$ is the orbit radius, $\psi$ is the angle of the photon with the orbit plane 
and $\gamma = \en/m_e = (1-v^2)^{-1/2} \ll 1$ is the electron Lorentz factor. 
In this paper we will assume that the orbit plane is horizontal.
Eq.(1) is obtained considering the photon as a classical pointlike particle 
moving along a definite light ray and comparing two expressions of the photon 
vertical angular momentum,
\begin{equation}
J_z =  R_{phot} \ \hbar \omega \ \cos\psi
\,,
\label{Jz}
\end{equation}
\begin{equation}
J_z = \hbar \omega \ \RB /v
\,.
\label{Jz'}
\end{equation}
The first expression is the classical one for a particle of horizontal momentum 
$k_h = \omega \cos\psi$ and impact parameter $R_{phot}$ with respect to the orbit axis.
The second expression comes from the invariance of the system \{electron + field\}
under the product of a time translation of $\Delta t$ times an azimuthal rotation 
of $\Delta \varphi = v \ \Delta t /\RB$. 

\begin{figure}[ht]
\includegraphics*[scale=0.8,clip,bb=70 547 411 748]{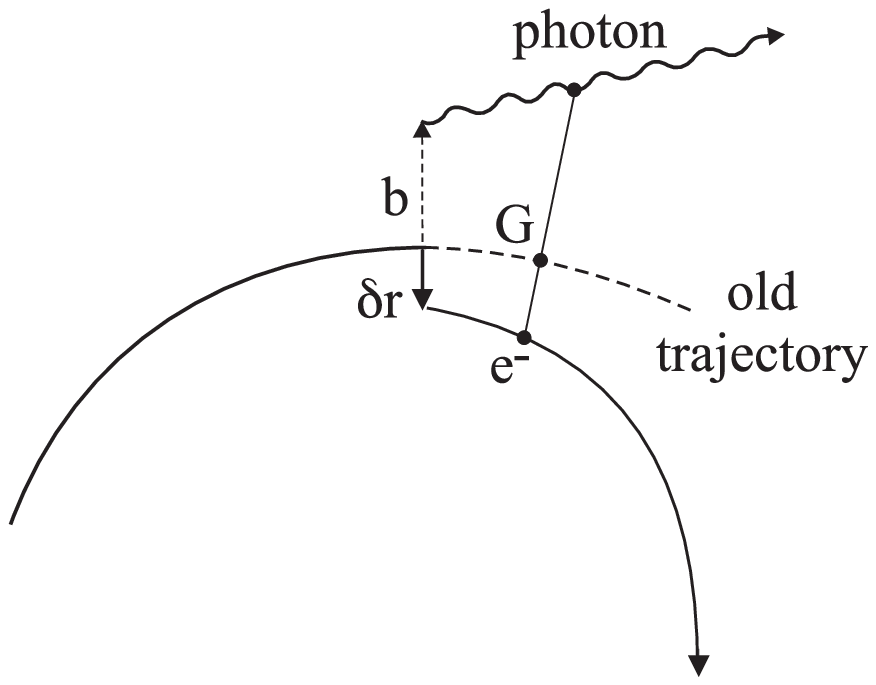}
\caption{Impact parameter $\bv$ of the photon and side-slipping $\deltare$ of the electron.}
\end{figure}

The impact parameter of the photon is connected, via angular momentum conservation, 
to a lateral displacement, or {\it side-slipping}, 
of the electron toward the center of curvature
\cite{Bignon,Qasmi}. The amplitude of this displacement is 
\begin{equation}
|\deltare| = { \hbar\omega \over\en - \hbar\omega } \ \bcl
\,,
\label{jump}
\end{equation}
so that the center-of-mass $G$ of the \{photon + final electron\} system continues the initial 
electron trajectory for some time, as pictured in Fig.1.
Whereas $\bcl$ is a classical quantity and can be large enough to be observed, 
$\deltare$ contains a factor $\hbar$ and is very small:
$ |\deltare| \sim (\omega / \omega_{c}) \; \lambdaC$, 
where $\lambdaC$ is the Compton wavelength 
and $\omega_{c} = \gamma^3/\RB$ the cutoff frequency.
Therefore the side-slipping of the electron is practically impossible to detect directly 
(in channeling radiation at high energy, 
it contributes to the fast decrease of the transverse energy \cite{cooling}).
However, in the classical limit $\hbar \to 0$, the number of emitted photons grows up
like $\hbar^{-1}$ and many small side-slips sum up
to a continuous lateral {\it drift velocity} $\delta \vv $ of the electron 
relative to the direction of the momentum: 
\begin{equation}
\delta \vv - {\pv\over \en} \equiv \vv = {2r_e \over3} \ \gamma \ {d^2 \Xv \over dt^2}
\,,
\label{drift}
\end{equation} 
$r_e = e^2/(4\pi m_e) $ being the classical electron radius%
\footnote
{we use rational electromagnetic equations, e.g.  div $\Ev = \rho \ $ instead of $4\pi \rho$. 
}. 
The distinction between $\vv$ and $\pv/\en$ is illustrated in Fig.2. 
Eq.(\ref{drift}) also results from a suitable definition of the electromagnetic 
part of the particle momentum \cite{Teitelboim,Qasmi}. 
The problematic {\it Schott term} $(2/3) r_e ({d^3 X^\mu / d\tau^3})$ 
of radiation damping can be interpreted \cite{Teitelboim} as the 
derivative of the drift velocity with respect to the proper time $\tau$. 
Thus the measurement of the photon impact parameter would constitute 
an indirect test of the side-slipping phenomenon and support
a physical interpretation of the Schott term.
\begin{figure}[ht]
\includegraphics*[scale=0.8,clip,bb=100 490 465 720]{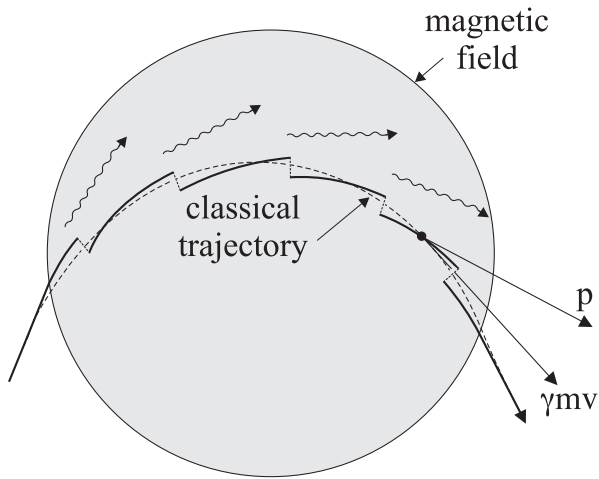}
\caption{Semi-classical picture of multiple photon emissions.
The dashed line represents the classical trajectory. The momentum $\pv$
is tangent to the semi-quantal trajectory and not equal to 
$\gamma m \, \vv$. The classical velocity $\vv$ is tangent to the 
dashed line.
}
\end{figure}

\section{Impact parameter profile in wave optics}

Using a sufficiently narrow electron beam, the photon impact parameter 
and its dependence on the vertical angle $\psi$ 
may be observed through an optical system such as in Fig.3. 
This system should be {\it astigmatic}, i.e. 
\begin{itemize}
\item{} horizontally focused on the transverse plane $P$, to see at which 
horizontal distance from the beam the radiation seems to originate.
\item{} vertically focused at infinity, to select a precise value of $\psi$.
\end{itemize}

\begin{figure}[ht]
\includegraphics{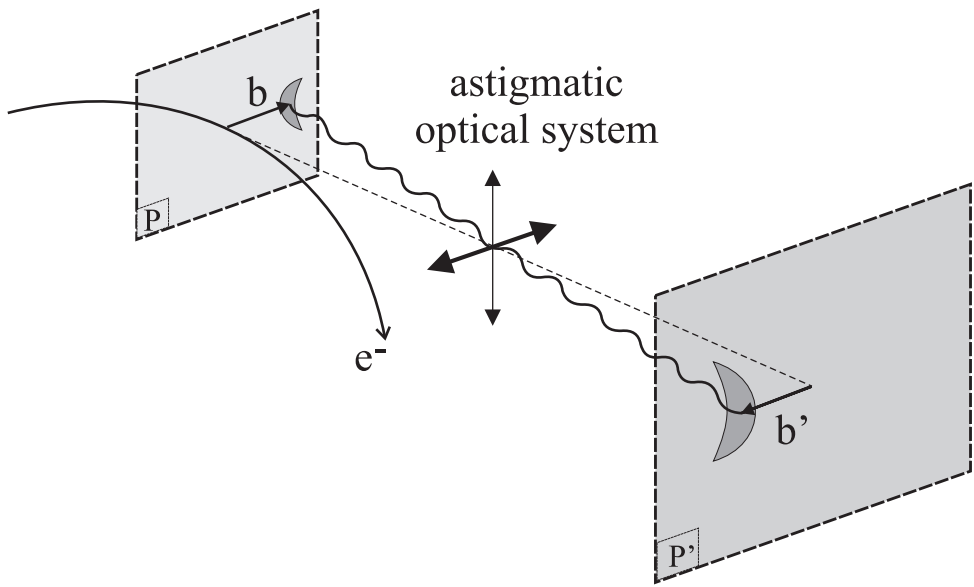}
\caption{Optical sytem for observing the horizontal impact parameter profile 
of synchrotron radiation at fixed vertical photon angle.
}
\end{figure}

\noindent
The horizontal projections of light rays emitted at three different times corresponding to 
electron positions $S_1, S, S_2$ are drawn in Fig.4. 
From what was said before, they are not tangent to the electron beam but 
to the dashed circle of radius $\RB+\bcl$ at, points $T_1, T, T_2$. 
Primed points are the images of unprimed ones by the lens.
$S_1$ and $S_2$ were taken symmetrical about the object plane $P$, 
such that the corresponding light rays come to the same point $M'$ of the image plane $P'$. 
When $S_1$ and $S_2$ are running on the orbit, $M'$ draws a classical {\it image spot} 
on plane $P'$ in the region $x' \ge \bcl'$, where $\bcl' = G \, \bcl$ 
and $G$ is the magnification factor of the optics, which from now on we will be taken 
equal to unity.
One can also say that the light rays do not converge to a point but 
form the {\it caustic} passing across $T'$. 
Classically, the intensity profile in the plane $P'$ behaves like $(x'- \bcl)^{-1/2}$.
Note that if synchrotron radiation was isotropic, emitted at zero impact parameter,
and if the optical system had a narrow diaphragm (for the purpose of increasing the depth-of-field),
the spot would be located at negative instead of positive $x'$.

\begin{figure}[ht]
\includegraphics{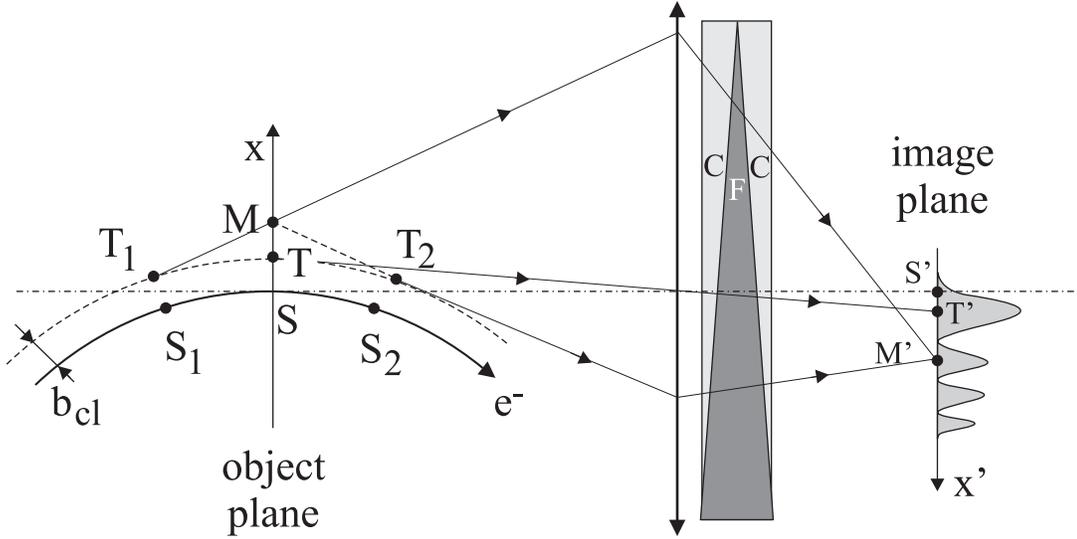}
\caption{Light rays of synchrotron radiation forming the image 
of the horizontal impact parameter profile. The profile intensity is 
schematically represented on the right side. "CFC" is a zero-angle dispersor.}
\end{figure}

Up to now we treated synchrotron radiation using geometrical optics. 
However this radiation is strongly self-collimated and the resulting 
self-diffraction effects must be treated in wave optics. 
For theoretical calculations, instead of the image spot in plane $P'$,  
it is simpler to consider its reciprocal image in plane $P$, which we call the {\it object spot}.
One must be aware that the latter is {\it virtual}, i.e. it does not represent
the actual field intensity in the neighbourhood of $S$. 
Its intensity is $| \Ev_{rad} |^2$ where
\begin{equation}
\Ev_{rad} = \Ev_{ret} - \Ev_{adv} 
\label{ret-adv}
\end{equation}
is the so-called {\it radiation field} and obeys the source-free Maxwell equations.
The distinction between the actual (retarded) field and $ \Ev_{rad} $ is illustrated
in Fig.5. 
%
\begin{figure}[ht]
\includegraphics*[scale=0.8,clip,bb=72 542 533 725]{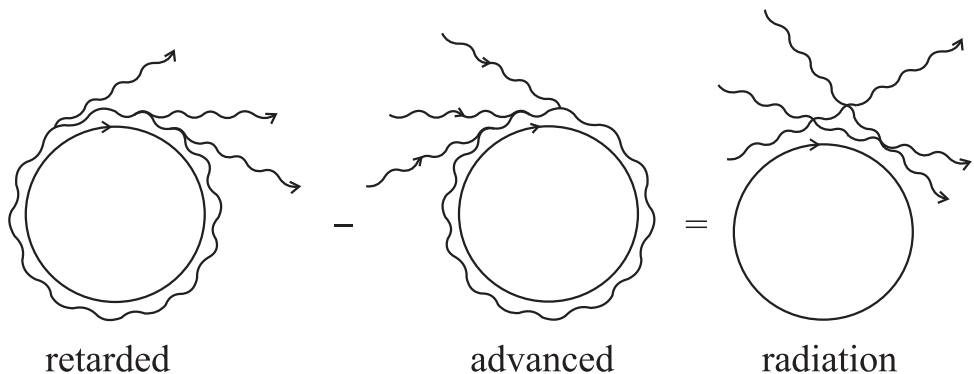}
\caption{Schematic representation of the relation beween the retarded, 
advanced and radiation fields}
\end{figure}

Using this point of view, one can say that the vertical cylinder of radius 
$R_{phot} = \RB + \bcl$ is the {\it caustic cylinder} of $ \Ev_{rad} $ 
(there is one such cylinder for each $\psi$). 
Thus the object spot amplitude can be taken from the known formula \cite{Landau} 
of the transverse profile of a wave near a caustic. 
At fixed frequency $\omega$ and vertical angle $\psi$ it gives
\begin{equation}
\hat\Ev_{rad} (\omega, x, \psi)
\propto {\rm Ai} \left( 2^{1/3} {\bcl(\psi) - x \over \BB }  \right)
\,,
\label{Landau-c}
\end{equation}
where Ai($\xi$) is the Airy function \cite{Abram}, 
$ \BB = \RB ^{1/3} \lambdabar ^{2/3} $ characterizes the width of the brillant region
of the caustic and $\lambdabar\equiv \lambda/(2\pi) = \omega^{-1}$.
The Airy function has an oscillating tail at positive $x-\bcl$ and 
is exponentially damped at negative $x-\bcl$. 
\begin{figure}[ht]
\includegraphics*[scale=1.0,clip,bb=-30 14 254 214]{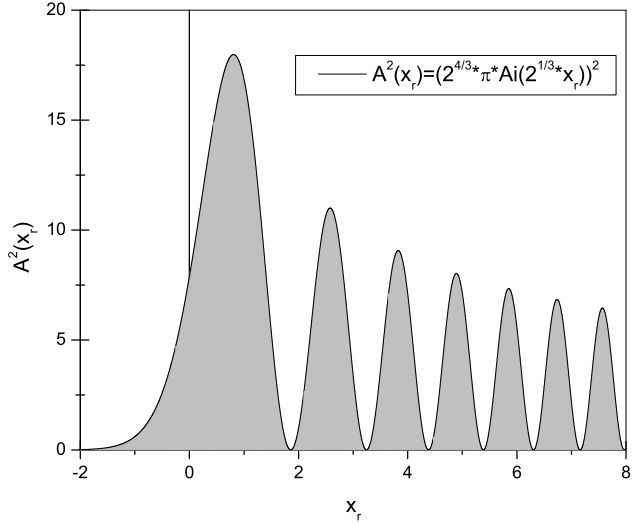}
\caption{Intensity of the horizontal impact parameter profile at fixed vertical angle $\psi$. 
The abscissa is $x_r = (x-\bcl) /\BB $, the ordinate is the value 
of $\Airi^2$ in Eq.(17).
}
\end{figure}
Fig.6 displays the intensity profile $|\hat\Ev_{rad}|^2$, in relative units, 
as a function of the dimensionless variable $x_r = (x-\bcl) /\BB $.
The oscillations can be understood semi-classically as an interference between light rays 
coming from symmetrical points like $T_1$ and $T_2$ in Fig.4. 
As said earlier these light rays come to the same point $M'$ of the image plane, 
therefore should interfere at this point. The phase difference between the two waves is 
\begin{equation}
2\delta =  \omega \ (t_2 - t_1) - k_{h} \ ( \, T_1M + MT_2 \,) 
\,,
\label{dephasage}
\end{equation}
where $t_{1,2}$ is the time when the electron is at $S_{1,2}$.
Denoting the azimuths of $S_1$ and $S_2$ by $-\varphi$ and $+\varphi$, we obtain
\begin{equation}
\delta = (\tan\varphi - \varphi) \ \omega \RB/v
\ \simeq \ \omega \RB \ \varphi^3/3
\ \simeq \ {1\over3}  \left( 2 {x - \bcl \over \BB}  \right)^{3/2} 
\,.
\label{2dephasage}
\end{equation}
This is in agreement, up to the residual phase $\pi/4$, with the 
the large $x$ behaviour in $x^{-1/4} \, \sin(\delta+\pi/4)$ 
of Eq.(\ref{Landau-c}). 

The result (\ref{Landau-c}) can also be obtained by recalling that the photon has a definite 
angular momentum $J_z = \hbar \omega \ \RB /v$ about the orbit axis. 
For fixed vertical momentum $\hbar k_z = \hbar\omega\psi$, 
we have the following radial wave equation, using cylindrical coordinates $(\rho, \varphi, z)$:
\begin{equation}
\left[ {\partial^2 \over d\rho^2} 
+ \rho^{-1} {\partial \over d\rho} 
\omega^2 - k_z^2 -  { J_z^2 \over \rho^2 }
\right] \ \Ev = 0
\,,
\label{RADIAL}
\end{equation}
We have neglected the photon spin and approximated the centrifugal term $ L_z^2 / \rho^2$
by $ J_z^2 / \rho^2$. 
In the vicinity of the classical turning point $\rho = \RB + \bcl$, 
we can use a linear approximation of the centrifugal potential 
and neglect the term $\rho^{-1} {\partial / d\rho} $.
This lead us to the Airy differential equation with the argument of (\ref{Landau-c}). 

The profile function can also be calculated in a standard way.
The electric radiation field $\Ev_{rad}$ can be expanded in plane waves as
\begin{equation}
\Ev_{rad}(t,\rv) = \int { d^3\kv \over (2\pi)^3 } \
\Re \left\{   \tilde\Ev(\kv) \ e^{i\kv\cdot \rv - i\omega t } \right\}
\,,
\label{EXPANSION}
\end{equation}
with $\omega=|\kv|$. The momentum-space amplitude%
\footnote{
From now on we omit the subscript {\it "rad"} of $\Ev$.
}
is given by
\begin{equation}
\tilde\Ev(\kv) =  e \ \int_{-\infty}^\infty 
dt_e \ \vv_{\perp}(t_e) \ \exp\left[i\omega t_e - i\kv\cdot \rv_e(t_e)\right]
\,,
\label{K-AMPLITUDE}
\end{equation} 
where $\vv_\perp$ is the velocity component orthogonal to $\kv$. 
The electron trajectory is parametrized as 
\begin{equation}
\rv_e(t_e) = \left( \RB \cos \varphi - \RB \ ,
\RB \sin \varphi \ , 0 \right)
\,, \qquad 
\varphi = v \, t_e / \RB 
\label{TRAJ}
\end{equation}
The energy carried by $\Ev_{rad}$ through a strip $[x \,, x+dx]$ of plane $P$,
in the frequency range $ [\omega , \omega + d\omega] $
and in the vertical momentum range $[k_z \,, k_z + dk_z]$ is 
\begin{equation}
dW = 2 dx \ { d\omega \ dk_z \over (2\pi)^2 } \ 
| \hat\Ev(\omega,x,\psi) |^2 
\,,
\label{STRIP}
\end{equation}
where
$\hat\Ev(\omega,x,\psi)$ is the partial Fourier transform 
\begin{equation}
\hat\Ev(\omega,x,\psi) = 
\int { dk_x \over 2\pi } \ e^{i k_x \; x} \ \tilde\Ev(\kv) 
\label{HAT}
\end{equation}
evaluated at $k_y \simeq \omega$ and $k_z \simeq \omega \; \psi $.
The result of Eqs.(\ref{K-AMPLITUDE} - \ref{HAT}) is 
\begin{equation}
{dW \over d(\hbar\omega) \ dx \ dk_z}
= {\alpha \over 8\pi^3} \ \Airi^2 \left( {\bcl - x \over \BB}  \right)
\left[ \Airi'^2(u) +  \left( {\psi\over\theta_0} \right)^2 \Airi^2(u)  \right]
\,,
\label{PROFIL}
\end{equation}
\begin{equation}
= {1 \over 2\pi} \ \Airi^2 \left( {\bcl - x \over \BB}  \right)
{dW \over d(\hbar\omega) \ (d\Omega/\theta_0^2)}
\,,
\label{PROFIL'}
\end{equation}
where $\alpha = e^2/(4\pi\hbar) = 1/137$, 
\begin{equation}
u \equiv s^2 \ (1+\gamma^2 \psi^2) / 2 \ = \ \bcl  / \BB
\,,\quad
s \equiv (\omega /\omega_c)^{1/3} 
\,,\quad 
\omega_c = \gamma^3/\RB 
\,, 
\label{S}
\end{equation}
\begin{equation}
\BB \equiv \RB^{1/3} \lambdabar^{2/3} \ = s \gamma \lambdabar 
\,,\qquad
\theta_0 \equiv (\lambdabar / \RB)^{1/3} = (s \gamma)^{-1}
\,
\label{theta0}
\end{equation}
and $\Airi(u)$ is a re-scaled Airy function:
\begin{equation}
\Airi (u) = 2^{4/3} \pi \ Ai \left( 2^{1/3} u \right) 
\,.
\label{A}
\end{equation}
Expression (\ref{PROFIL'}) relates the 
$(x,\psi)$ profile to the standard the angular distribution. 
The two terms of the square braket of (\ref{PROFIL}) correspond to the horizontal 
and vertical polarizations respectively. 
\begin{figure}[ht]
\includegraphics*[scale=1.2,clip,bb=0 14 320 217]{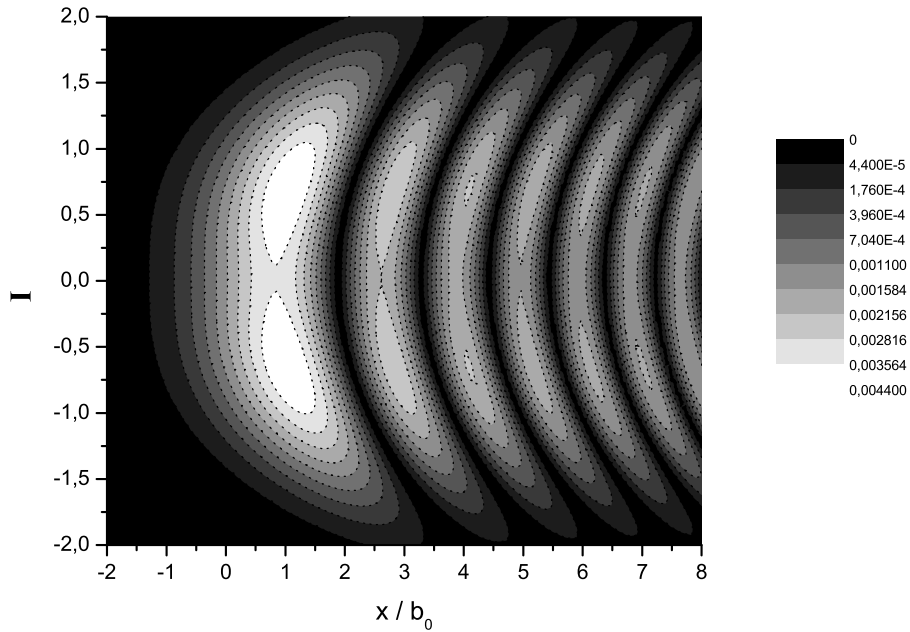}
\caption{$(x,\psi)$ profile of synchrotron radiation (Eqs.16-17), 
in the limit $s^3 = \omega /\omega_c \ll 1$. The $i^{th}$ level curve corresponds 
to the fraction $(i/10)^2$ of the maximal intensity.}
\end{figure}

An example of the $(x, \psi)$ profile is shown in Fig.7. We recall that it is 
obtained using an astigmatic lens. Therefore it differs from the standard $(x,z)$ profile
\cite{Hofmann,Bosch,Chubar} formed by a stigmatic lens. 
The number of photons per electron and per unit of $\ln \lambda$ in 
the first bright fringe is, in the $s \ll 1$ limit, 
\begin{equation}
{dN_{phot} \over d \lambda / \lambda } \simeq {1 \over 70}
\,.
\label{Nphot}
\end{equation}
The second fringe contains 3 times less photons.

\section{Applications}

Although the impact parameter cannot be sharply defined in wave optics, 
Eq.(\ref{PROFIL}) keeps a trace of the classical prediction (\ref{bcl}):
as $\gamma$ or $\psi$ is varied, the $x$-profile
translates itself as a whole, comoving with the classical point $x=\bcl$. 
The observation of this feature would constitute an indirect test of the phenomenon
of electron side-slipping.
The $\psi^2$ dependence of this translation is responsible for the curvature of the fringes.
The $\gamma^{-2}$ dependence may be more difficult to observe: 
$\bcl$ should be as large as possible compared to the horizontal width of the 
electron beam, one one hand, and to the FWHM width of the first
fringe, $\simeq 1.3 \ \BB$, on the other hand.
Since the profile intensity (\ref{PROFIL}) decreases very fast at large $u=\bcl/\BB$,
a sensitive detector is needed. 

The following table summarizes the various length scales which appear in 
synchrotron radiation.

\[
\begin{array} {| l | c | c | c | c |}
\hline 
\hbox{ bending } & & & & \\
\hbox{ radius } & \RB & & & \\
\hline 
\hbox{ longitudinal } & & & & \\
\hbox{ distances } & 
\LLm = \RB \ \gamma^{-1} &  \LL = \RB^{2/3} \ \lambdabar^{1/3} & & \LLp = \gamma^2 \lambdabar \\
\hline 
\hbox{ transverse } & & & & \\
\hbox{ distances } & 
\BBm = \RB \ \gamma^{-2} &  & \BB = \RB^{1/3} \ \lambdabar^{2/3} & \BBp = \gamma \lambdabar \\
\hline 
& & & & \\
\hbox{ wavelengths } & 
\lambdabarc = \RB \ \gamma^{-3} &  &  &  \lambdabar \\
\hline 
\end{array} 
\]
The four quantities of a given row or line are in geometric progression 
of ratio $ \gamma^{-1} $ or $ s^{-1} $ respectively. Going along (or parallel to) the diagonal, 
the ratio is $\theta_0 = (s \gamma)^{-1}$.
The subscripts of the different $l$'s and $b$'s are the powers of $\gamma$. 

Synchrotron radiation is not emitted instantaneously, 
but while the electron runs within a distance $l_f$, called {\it formation length}, 
from point $S_i$. 
Thus $2 l_f$ is a minimum length of the bending magnet.
A conservative estimate of $l_f$ may be 
\begin{equation}
l_f = {\rm Max} \{2\LLm, 3\LL \}
\,.
\label{lf}
\end{equation}
In addition, the $m^{th}$ fringe of the $(x,\psi)$ profile comes from points 
$S_1$ and $S_2$ at distance 
\begin{equation}
l^{(m)} = (3m\pi)^{1/3} \LL 
\label{lfringe}
\end{equation}
from point $S$. To observe it, the magnet half-length should therefore be larger than
$l_{min} = l_f + l^{(m)}$. 
The distance between the object plane and the lens should be larger than this $l_{min}$,
plus a few $\LLm$ so that the lens can accept the ray comming from $T_2$ but not intercept the 
near field. 

Some numerical examples are given in the following Table. 
\[
\begin{array} {| c | c | c | c |c |}
\hline 
\gamma & 200 & 200 & 6000 & 1000  \\
\hline 
\RB & 0.8\ {\rm m} & 80\ {\rm m} & 4000\ {\rm m} 
& 3\ {\rm m} \\
\hline 
\LLm = \RB\ \gamma^{-1} & 4\ {\rm mm} & 0.4\ {\rm m} & 0.67\ {\rm m} 
& 3\ {\rm mm} \\
\hline 
\BBm = \RB\ \gamma^{-2} & 20\ \mu{\rm m} & 2\ {\rm mm} & 0.11\ {\rm mm} 
& 3\ \mu{\rm m} \\
\hline 
\lambdabarc = \RB \ \gamma^{-3} & 0.1\ \mu{\rm m} & 10\ \mu{\rm m} & 19\ {\rm nm} 
& 3\ {\rm nm} \\
\hline 
\lambdabar & 0.1\ \mu{\rm m} & 0.1\ \mu{\rm m} & 0.1\ \mu{\rm m} 
& 3\ \mu{\rm m} \\
\multicolumn{1}{|c}{} &
\multicolumn{3}{|c|}{\hbox{(visible domain)}} &
\multicolumn{1}{c|}{\rm (infrared)} 
\\
\hline 
s = (\lambdabar/\lambdabarc)^{-1/3} & 1 & 4.6 & 0.57 
& 0.1 \\
\hline 
\theta_0 = s^{-1} \gamma^{-1} & 5\ {\rm mrad} & 1.1\ {\rm mrad}  & 0.29\ {\rm mrad} 
& 10\ {\rm mrad} \\
\hline 
\LL = s^{-1} \ \LLm & 4\ {\rm mm} & 86\ {\rm mm} & 1.2\ {\rm m} 
& 30\ {\rm mm} \\
\hline 
\BB = s^{-2} \ \BBm & 20\ \mu{\rm m} & 93\ \mu{\rm m} & 0.34\ {\rm mm} 
& 0.3\ {\rm mm} \\
\hline 
\end{array} 
\]

A practical application of the fringe pattern of Fig.7 is the measurement 
of an horizontal beam width. Since the successive fringes are more and more dense,
they can probe more and more smaller widthes. For a gaussian beam of r.m.s. width $\sigma_x$, 
the contrast between the $m^{th}$ minimum and the $m+1^{th}$ maximum is
\begin{equation}
a_m 
= \exp\left[-2 (3\pi m)^{2/3} \ (\sigma_x / \BB)^2  \right]
\,.
\label{CONTRAST}
\end{equation}
The vertical beam size and the horizontal angular divergence have practically no blurring effect on 
the observed profile. 
This is not the case for the vertical beam divergence. However, as long as this divergence
is small in units of $\theta_0$, the blurring is small. 

The $x$ scale parameter $\BB$ grows with $\lambda$. Therefore a too large 
passing band of the detector will blur the fringes. For instance the contrast
of the $m^{th} $ fringe is attenuated by a factor $2/\pi$ when the relative passing
band is $\Delta\lambda/\lambda = 1/(2m)$.
It is possible to restore a good contrast for a few successive fringes
by inserting a dispersive prism ("CFC" in Fig.4) at some distance before the image plane,
such that the $\lambda$-dependent deviation by the prism 
compensates the drift of the $m^{th} $ fringe. This allows to increase the 
passing band, hence the collected light, without loosing resolution.

\section{Conclusion}

\medskip

The analysis of synchrotron radiation simultaneously in horizontal impact parameter $x$
and vertical angle $\psi$, which, to our knowledge, has not yet been done, can open the way
to a new method of beam diagnostics. Only simple optics elements are needed. 
There is no degradation of the beam emittence 
(contrary to Optical Transition Radiation) 
and no space charge effect at high beam current 
(such effects may occur with Diffraction Radiation).
In addition, the observation of the curved shape of the fringes and the precise measurement 
of their distances to the beam would give an indirect support
to the phenomenon of electron side-slipping and to a physical interpretation of the 
Schott term of radiation damping.


\begin{thebibliography}{99}
\bibitem{Bignon} 
X. Artru, G. Bignon,
Electron-Photon Interactions in Dense Media, H. Wiedemann (ed.), 
NATO Science series II, vol. 49 (2002), pages 85-90. 
\bibitem{Qasmi} 
X. Artru, G. Bignon, T. Qasmi,
Problems of Atomic Science and Technology 6 (2001) 98 ; arXiv:physics/0208005.
\bibitem{cooling} 
X. Artru, Phys. Lett. A 128 (1988) 302, Eqs.(15-16). 
\bibitem{Teitelboim} 
C. Teitelboim, Phys. Rev. D1 (1969) 1572; D2 (1970) 1763.
See also C. L\'opez and D. Villarroel, Phys. Rev. D11 (1975) 2724.
I thank Prof. V. Bordovitsyn for pointing me these references.
\bibitem{Abram} 
M. Abramowitz and I. Stegun {\it Handbook of Mathematical Functions}, Dover, 1970.
\bibitem{Landau} 
L. Landau and E. Lifshitz, {\it The Classical Theory of Fields}, §7.7 (Addison-Wesley 1951).
\bibitem{Hofmann} 
A. Hofmann, F. M\'eot, Nucl. Inst. Meth. 203 (1982) 483. 
\bibitem{Bosch}
R.A. Bosch, Nucl. Inst. Meth. in Physs. Res. A 431 (1999) 320. 
\bibitem{Chubar} 
O. Chubar, P. Elleaume, A. Snirigev, Nucl. Inst. Meth. in Physs. Res. A 435 (1999) 495. 
\end{thebibliography}
\end{document}